\documentclass{elsart}
\usepackage{graphics}

\usepackage{color}
\usepackage{epsfig}
\usepackage{graphics}
\usepackage{psfrag}
\usepackage{lscape}

\begin{document}
\begin{frontmatter}
\title{Microcanonical scaling in small systems}

\author[ER]{M. Pleimling\thanksref{erl}},
\author[ER]{H. Behringer} and
\author[ER]{A. H\"{u}ller}
\address[ER]{Institut f\"ur Theoretische Physik 1, Universit\"at Erlangen-N\"urnberg, D--91058 Erlangen, Germany}
\thanks[erl]{Corresponding author:
tel: +49-9131-8528451; fax: +49-9131-8528444;
e-mail: pleim@theorie1.physik.uni-erlangen.de}

\begin{abstract}
A microcanonical finite-size scaling ansatz is discussed. It exploits the existence
of a well-defined transition point for systems of finite size in the microcanonical ensemble.
The best data collapse obtained for small systems yields values for the critical exponents
in good agreement with other approaches.
The exact location of the infinite system critical point is not needed
when extracting critical exponents from the microcanonical finite-size scaling theory.
\end{abstract}

\end{frontmatter}

\newpage

In the canonical ensemble phase transitions appear exclusively in infinite
systems. At the critical point of a continuous phase transition denoted by
$T_c$ and $h_c$ the Gibbs
free energy as a function of the temperature $T$ and of the field $h$
conjugate to the order parameter develops a non-analytic behaviour. As a
consequence some thermodynamic functions as e.\,g. the specific heat and the
susceptibility exhibit power law singularities $|T-T_c|^{-\alpha}$ and
$|T-T_c|^{-\gamma}$ in the vicinity of $T_c$ and $h=h_c$ fixed, commonly with
non-classical critical exponents $\alpha$ and $\gamma$. 
In finite systems all these singularities are rounded in the canonical ensemble.
The appearence of this rounding is taken into account by finite-size scaling
theory originally proposed on phenomenological grounds
\cite{Bar83,Fis72,Pri84}. In the asymptotic 
limit $L \to \infty$ and $T \to T_c$ the behaviour of finite-size
quantities is governed by scaling functions. The scaling
functions are basically determined by the ratio $L/\xi(T)$ with $\xi(T)$
being the correlation length and $L$ the linear extension
of the system. Note that the validity of finite-size scaling relations has been
proven in renormalization group theory \cite{Bar83}. Over the years finite-size scaling
theory has been shown to be
a valuable tool for extracting critical exponents from finite-size data, as
obtained for example from numerical simulations.

In the microcanonical analysis of finite systems one considers the entropy
$S_N(E,M) = \ln \Omega_N(E,M)$ as a function of the energy
$E$ and of the magnetization $M$ \cite{Gro01}. The density of states of a
system with $N$ spins is denoted by $\Omega_N(E,M)$ and natural units with $k_B
=1$ are chosen. The entropy surface exhibits a well-defined transition
point at an energy $E_c$ and magnetization $M_c$
although in finite systems $S_N(E,M)$ is everywhere perfectly analytic.
The microcanonical analysis also 
shows that typical features of symmetry breaking, as for example the abrupt
onset of several order parameter branches when the transition point is crossed from above, are
already encountered in small systems \cite{Kas00,Hue01,Behr04}. With regard to these intriguing effects, it is tempting
to ask whether a direct analysis of the microcanonical entropy also allows 
the determination of critical quantities from finite-size data.

In this Letter we provide numerical evidence that it is in principle
possible to extract critical exponents from the behaviour of finite
microcanonical systems. Considering different classical spin models
belonging to different universality classes we show that the critical exponents can
be obtained from systems which are moderately small. The scaling ansatz to
be discussed in the following does not need any {\it a priori}
knowledge of the infinite system. Especially, the knowledge of the exact location of the critical
point of the infinite system is not needed.

In the following we discuss the nearest neighbour ferromagnetic Ising (I)
model in three dimensions with the Hamiltonian
\begin{equation} \label{gl1}
{\mathcal H}_I = - J \, \sum\limits_{\langle i,j \rangle} \, S_i \, S_j
\end{equation}
as well as a generalized Ising (GI) model with equivalent nearest and next-nearest neighbour
interactions defined by the Hamiltonian
\begin{equation} \label{gl2}
{\mathcal H}_{GI} = - J \, \sum\limits_{\langle i,j \rangle} \, S_i \, S_j
- J  \, \sum\limits_{( i,k )} S_i \, S_k.
\end{equation}
Here, $J > 0$ is the coupling constant and the spins $S_i$ can take on the
values $\pm1$. The first sum in equation (\ref{gl2}) extends over nearest
neighbour bonds whereas the second sum is over bonds connecting next-nearest neighbours.
On the simple cubic lattice, a spin has 
6 nearest neighbours and 12 next-nearest neighbours which yields the ground state
energy per spin $\varepsilon_{GI} = -9 \, J$ for the GI model. On the same lattice the simple Ising model has
the ground state energy $\varepsilon_I= -3 \, J$. The continuous phase transitions observed in both models
belong to the same universality class. In addition, we investigate the three
state Potts model in two dimensions whose Hamiltonian reads
\begin{equation} \label{potts}
{\mathcal H}_P = - J \, \sum\limits_{\langle i,j \rangle} \delta_{ S_i,S_j},
\end{equation}
with the spins taking on  the values $S_i = 1,2,3$.

In a microcanonical analysis of finite systems with $N = L^d$ spins ($d$ being the number of space
dimensions), the object of interest is the density of states $\Omega_N(E,M)$ as a function of
the energy $E$ and the magnetization $M$. The spontaneous magnetization $M_{sp,N}(E)$
is defined to be the value of $M$ where the entropy $S_N(E,M)$ at a fixed
value of the energy $E$ has its maximum with respect to $M$:
\begin{equation}
M_{sp,N}(E) : \iff S_N(E, M_{sp,N}(E)) = \max_{M} S_N(E, M) \;.
\end{equation}
This definition of the microcanonical spontaneous magnetization assures
the equivalence of the canonical and microcanonical ensemble in the
thermodynamic limit.
At energies lower than a transition energy $E_{c,N}$ the entropy of the Ising
universality class exhibits two maxima at
$M = \pm M_{sp,N}$. When approaching  the finite-size transition point
$E_{c,N}$ from below the order parameter $M_{sp,N}$
vanishes with a square root behaviour in the finite system
\cite{Kas00,Behr04}. Thus the transition energy
$E_{c,N}$ can be localized with high precision. This vanishing is also reflected in the
divergence of the microcanonically defined finite-size zero-field
susceptibility \cite{Kas00}. In Figure 1
the variation of the spontaneous magnetization per spin, $m_{sp,N}=M_{sp,N}/N$, as a function of the
specific energy, $\varepsilon = E/N$, is shown for the two-dimensional
nearest neighbour Ising model on an infinite
and on a finite square lattice with $32^2$ spins. Both curves coincide at low energies. At higher
energies, the square root behaviour close to the finite-size transition point
can be observed. A similar behaviour is found for the entropy
surface of finite two-dimensional three state Potts models \cite{Beh03}.
Note that the discreteness of the
physical quantities of discrete spin systems such as the Ising model has to
be considered  with some care. There the
language used here refers to continuous functions that describe the discrete
data most suitably. In the microcanonical analysis of continuous
spin systems as e.g. the $XY$ \cite{Richt04} or the
Heisenberg model this concern does not exist.

\begin{figure}[h!]
\begin{center}
\epsfig{file=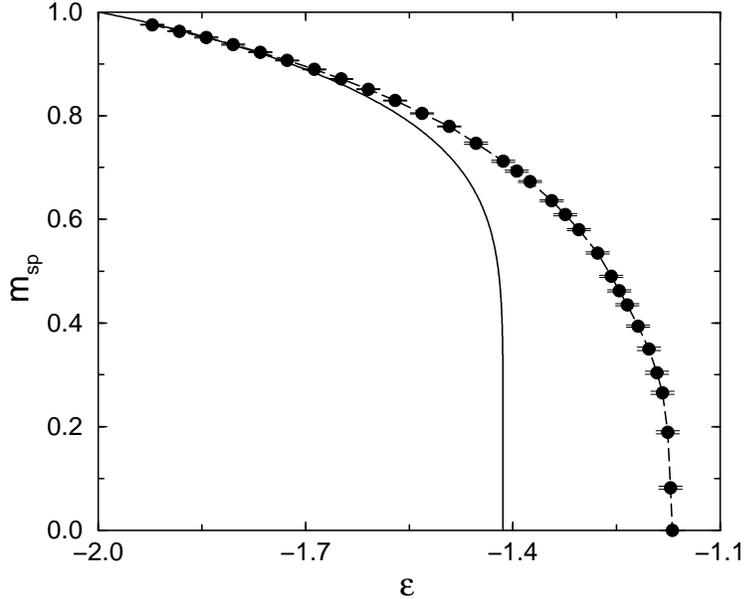,width=20em, angle=270}
\caption{
Microcanonically defined order parameter vs specific energy for two-dimensional Ising models
defined on an infinite square lattice (full line) and on a finite square lattice containing
$32 \times 32$ spins (symbols). The finite-size spontaneous magnetization vanishes at a well-defined
finite-size transition point.}
\end{center}
\end{figure}

In the canonical ensemble it is impossible to define an order parameter which 
exhibits the typical features of spontaneous symmetry breaking already in
finite systems. Usually, a pseudo critical temperature
is defined via the position of the maximum of some thermal quantity.
However, this definition introduces an ambiguity as different
quantities, e.g.\ the specific heat or the susceptibility, normally have their maxima located
at different temperatures.

In order to determine relevant microcanonical quantities with high precision, very accurate
estimations of the density of states are needed. 
The data presented here have been obtained with a very efficient algorithm \cite{Hue01}
based on the concept of transition observables \cite{Deo96}. From these observables derivatives of the
entropy with respect to the energy and/or magnetization are easily computed.
This is a point of 
utmost importance as it is not the entropy itself which enters into the microcanonical
analysis, but its derivatives with respect to $E$ and $M$.

The microcanonical finite-size scaling theory proposed and discussed in the following is formulated
in such a way as to take advantage of the existence of a well-defined
transition point in finite microcanonical  systems. 
This leads to the following scaling ansatz
\begin{equation} \label{gl3}
L^{\beta_\varepsilon/\nu_\varepsilon} \, m_{sp,N}\left(  \varepsilon_{c,N} -\varepsilon 
\right) \sim  W \left( C \, ( \varepsilon_{c,N} -\varepsilon ) \, L^{1/\nu_\varepsilon} 
\right)
\end{equation}
for the microcanonically defined order parameter in the limit of small scaling
variables $x = ( \varepsilon_{c,N} -\varepsilon ) \, L^{1/\nu_\varepsilon}$.
$C$ is a non-universal, model-dependent metric constant and $W$ is a universal scaling
function for a given universality class. Note that the microcanonical
critical exponents in equation (\ref{gl3}), which describe
the behaviour of various quantities with respect to the specific energy, are not identical to
the usual canonical exponents. For the order parameter one has $\beta_\varepsilon= \beta/(1- \alpha)$,
whereas the microcanonical correlation length critical exponent is given by $\nu_\varepsilon=
\nu/(1- \alpha)$ \cite{Pro95,Kas00}. Here $\beta$, $\alpha$, and $\nu$ are the canonical
critical exponents. As the microcanonical order parameter varies like a square
root in the vicinity of $\varepsilon_{c,N}$ for all finite system sizes $N$
\cite{Kas00,Behr04}
the scaling function $W$ is asymptotically given as a square root $W(x) \sim \sqrt{x}$
for small scaling variables $x$.
One remarkable feature of equation (\ref{gl3}) is the absence of any non-universal quantity related 
to the infinite system. Especially, the location of the infinite system critical point
$\varepsilon_{c, \infty}$ does not enter in the definition of the scaling
variable $x$.

\begin{figure}[h!]
\begin{center}
\epsfig{file=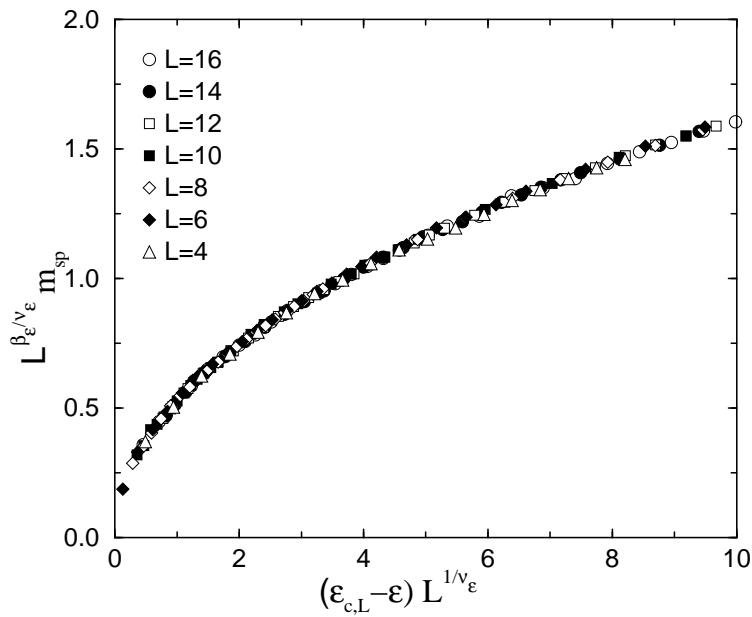,width=20em, angle=270}
\caption{
Microcanonical finite-size scaling plot for the three-dimensional Ising model. 
The values $\beta_\varepsilon/\nu_\varepsilon = 0.54 \pm 0.03$ and
$1/\nu_\varepsilon = 1.43 \pm 0.04$ result from the best data collapse.}
\end{center}
\end{figure}

In Figure 2 we test the ansatz (\ref{gl3}) by plotting $L^{\beta_\varepsilon/\nu_\varepsilon} 
\, m_{sp,N}$ as a function of $x = ( \varepsilon_{c,N} -\varepsilon ) \, L^{1/\nu_\varepsilon}$
for several, altogether rather small, three-dimensional Ising models. Here, $L$ ranges from 4 to 16. 
Using a recently proposed method for quantifying the nature of a data collapse \cite{Bha01},
the optimal exponents and their error bars can be obtained. For the 3d Ising model our small system data yield
the values $\beta_\varepsilon/\nu_\varepsilon = 0.54 \pm 0.03$ and
$1/\nu_\varepsilon = 1.43 \pm 0.04$, in remarkable agreement with the expected values
$\beta_\varepsilon/\nu_\varepsilon = 0.52$ and $1/\nu_\varepsilon = 1.43$. For the GI model,
a similar study for system sizes ranging from $L=4$ to $L=12$ yields the values
$\beta_\varepsilon/\nu_\varepsilon = 0.51 \pm 0.03$ and
$1/\nu_\varepsilon = 1.43 \pm 0.04$, which  again agrees with the literature 
values. For the 2d three state Potts models system sizes ranging from $L=6$
to $18$ have been investigated. The best data collapse results in the
estimates $1/\nu_\varepsilon = 0.79 \pm 0.03$ and
$\beta_\varepsilon/\nu_\varepsilon = 0.13 \pm 0.03$ for the critical exponents. The agreement with the
exactly known values  $1/\nu_\varepsilon = 4/5$ and
$\beta_\varepsilon/\nu_\varepsilon = 2/15$ is again very good considering
the smallness of the systems.

\begin{figure}[h!]
\begin{center}
\epsfig{file=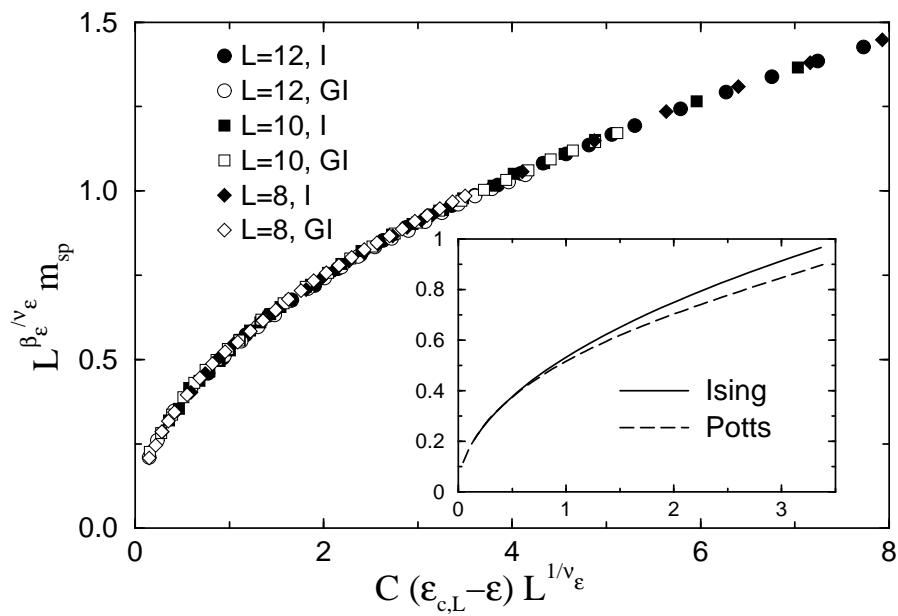,width=20em, angle=270}
\caption{
Microcanonical finite-size scaling plot both for the Ising (I) and for the GI model.
Adjusting the non-universal metric constants, see equation (\ref{gl3}), with $C_{GI}/C_{I}= 0.219$,
the data points of both models fall on a unique master curve, thus demonstrating the universality of
the finite-size scaling function $W$. The inset shows the scaling functions
for the Ising model (solid curve) and the 2d Potts model (dashed). The scaling variable $x$ of the Potts
model is rescaled so that the amplitude of the square root function in the
limit $x\to 0$ are identical (compare relation (\ref{gl3})). The curves are
only displayed for the range of the  scaling variable for which data for the Potts
model have been
obtained.}
\end{center}
\end{figure}

In Figure 3 we investigate whether the scaling function $W$ is indeed
universal. Adjusting the
values of the non-universal constants $C_{I}$ and $C_{GI}$, the data of both models should fall on a common master
curve for models within the same universality class. With  $C_{GI}/C_{I}= 0.219$ a unique curve is observed,
thus demonstrating the universality of the finite-size scaling function $W$.
In the inset of Figure 3 the scaling functions of the Ising model and of the
Potts model are compared. As the two model systems belong to different
universality classes different scaling functions are expected, which is
indeed confirmed by the data.

Our approach differs from earlier attempts at a microcanonical finite-size
scaling theory \cite{Kas00,Bru99,Kas01,Hov04,Des88}
in various regards. In Ref.\ \cite{Kas00,Kas01} it was supposed that the entropy of finite
systems was a homogeneous function in the vicinity of the transition point
$\varepsilon_{c,\infty}$ of the infinite system.  The resulting scaling
relations involve $\varepsilon_{c,\infty}$ in the scaling variable and lead
to rather poor results for the range of system sizes considered in the
present work. However, the scaling relations deduced in \cite{Kas00} are
expected to be valid in the asymptotic regime $L \to \infty$. The authors of Ref.\ \cite{Des88}
studied the enthalpy instead of the entropy and developed a finite-size scaling theory
in complete analogy to the canonical case. 
In the same way as in the canonical ensemble 
they defined a
finite-size order parameter which differs from zero for all energies. No noticeable differences
with canonical results were found in that approach.

In conclusion, we have presented numerical evidence that 
a suitable microcanonical finite-size scaling ansatz with the correct values of the critical exponents
leads to good data collapse of
microcanonical finite-size quantities of moderately small systems. It follows from this microcanonical ansatz that critical exponents
can easily be obtained from finite microcanonical systems without any {\it a priori} knowledge of
infinite system quantities. The expected universality of the scaling
function for a given universality class has been demonstrated. The
microcanonical finite-size scaling theory, presented for the order parameter
in the present work, can also be formulated more generally in terms of scaling relations of
the entropy of finite systems considered as a
function of the energy, the magnetization and the inverse system size \cite{Beh_diss04}. 


\end{document}